# A CMUT-Based Transcranial Focused Ultrasound Platform for Blood-Brain Barrier Opening in Small Animal Models

M. Sait Kilinc,[1] Reza Pakdaman Zangabad,[2] Victor Menezes,[2] Hohyun Lee,[2] Costas Arvanitis,[2,3] and F. Levent Degertekin [1,2,4,5,6*]

[1]School of Electrical and Computer Engineering, Georgia Institute of Technology, Atlanta, Georgia, United States

[2]Woodruf School of Mechanical Engineering, Georgia Institute of Technology, Atlanta, Georgia, United States

[3]Coulter Department of Biomedical Engineering, Georgia Institute of Technology and Emory University, Atlanta, Georgia, United States

[4]801 Ferst Drive NW, Atlanta, Georgia, United States 30332

[5]Senior author

[6]Lead contact

*Correspondence: levent.degertekin@me.gatech.edu

## SUMMARY

Drug delivery to the brain is limited by the blood-brain barrier (BBB). We developed a capacitive micromachined ultrasonic transducer (CMUT)-based transcranial focused ultrasound system capable of both delivering therapy via BBB opening and monitoring microbubble activity across a broad frequency range. The performance of the geometrically focused half-ring array consisting of five transmitters and one receiving element was first assessed through simulations and *in-vitro* acoustic measurements with microbubbles. Use of phase-inversion (PI) during transmission effectively suppressed CMUT-generated harmonics and enhanced broadband detection of microbubble emissions. In rats, the same system achieved spatially localized BBB opening, confirmed by T1-weighted magnetic resonance imaging. BBB permeability mapping using dynamic contrast-enhanced magnetic resonance imaging ($K_{trans}$) scaled with pressure. Time-resolved acoustic spectra captured microbubble arrival and decay kinetics, and 7-20dB enhancement in the effective dynamic range is observed with PI processing of acoustic emission signals. Together, these findings establish an integrated CMUT platform for combined therapeutic and sensing applications for BBB opening in small animal models, providing a foundation for future real-time, frequency-agile, closed-loop control of ultrasound-mediated drug delivery to the brain.

## INTRODUCTION

Neurological disorders and diseases, such as Alzheimer's, Parkinson's, and brain cancer remain challenging pharmacologic targets in part because of the restrictive nature of the blood-brain barrier (BBB) [1-3]. Transcranial focused ultrasound (tFUS) in combination with intravenously administered microbubbles (MBs) has emerged as a promising method for transient, localized, and noninvasive BBB disruption for targeted drug delivery to the brain [1, 4-11]. The overall treatment workflow for MB-assisted BBB opening is illustrated in Figure 1A. The procedure begins with intravenous MB administration, followed by targeted sonication through skull while real-time passive monitoring of acoustic emission is employed for adjusting sonication level to maintain stable MB oscillations and minimize the risk of inertial cavitation. This workflow inherently involves two simultaneous processes comprising therapeutic ultrasound transmission for activating MBs and passively listening to MB acoustic emissions [6]. The transmit frequency typically ranges from 0.2 to 1 MHz in clinical systems [5, 12], while preclinical systems may operate at higher frequencies, up to ~3 MHz [13, 14], depending on subject size and MB characteristics. The receive bandwidth commonly spans 0.1 to 6 MHz to capture subharmonic ($f_0/2$), harmonic ($n*f_0$), and ultraharmonic ($(n+½)*f_0$) components of acoustic emissions (Fig. 1B) with adequate signal-to-noise ratio for monitoring or feedback control purposes [6]. Because conventional tFUS systems are based on piezoelectric transducers that are intrinsically narrowband, they lack the flexibility in operating frequency and often require separate transmitter and receiver transducers to effectively detect the different spectral components [15, 16]. This adds to the system complexity and introduces monitoring challenges due to spatial mismatch in the field of view between the transmit and receive systems [4, 8, 12, 14, 17, 18].

In recognition of the above challenges researchers have started assessing the performance and abilities of capacitive micromachined ultrasonic transducers (CMUTs) for MB-assisted BBB opening [19, 20]. Properties like the wide and tunable bandwidth [21-23] (a typical CMUT frequency response shown in Fig. 1B), high receive sensitivity [24], and integration with on-chip electronics [25, 26] make CMUTs suitable receivers for acoustic emission detection. Consequently, several recent studies have demonstrated their use to passive cavitation detection (PCD) in MB-assisted BBB-opening studies [19, 20, 27], while their integration with low-noise transimpedance amplifiers (TIAs) has allowed to capture broadband acoustic emission signals with high sensitivity [27]. Other studies have employed narrow acquisition windows to isolate transient high-amplitude emissions and improve the detection of inertial cavitation [20]. The use of CMUTs as transmitters for BBB opening has received less attention due to their inherent electromechanical nonlinearity [28], as detection of harmonics is critical for acoustic emission based MB monitoring [13, 15, 18]. As a potential solution to this issue, in our earlier work we proposed using polyphase modulation (PM) [29], in which CMUTs are driven with coded and phase-shifted pulses to suppress device-generated harmonics [30], creating a path to reconfigure CMUT technology for an all-CMUT based system for BBB opening applications.

Here, we present a fully CMUT-based tFUS system with integrated transmit/receive capability and evaluate its potential for MB-assisted BBB opening in small-animal models. Building upon our prior single-element CMUT studies [29], we designed and fabricated a geometrically focused array comprising five transmitting elements and one receiving element on a cylindrical surface (Fig. 1C) To evaluate feasibility, we performed *in vitro* experiments to assess and fine tune the CMUT tFUS system. Subsequently, we conducted *in vivo* BBB-opening experiments in rats and assessed the potential of phase inversion (PI) based signal processing to monitor the MB acoustic emissions for potential closed loop operation. The CMUT-based platform combining therapeutic ultrasound transmission and broadband, sensitive acoustic emission detection represents a critical step toward engineering wideband tFUS systems capable of real-time feedback control MB-assisted BBB opening for targeted drug-delivery applications.

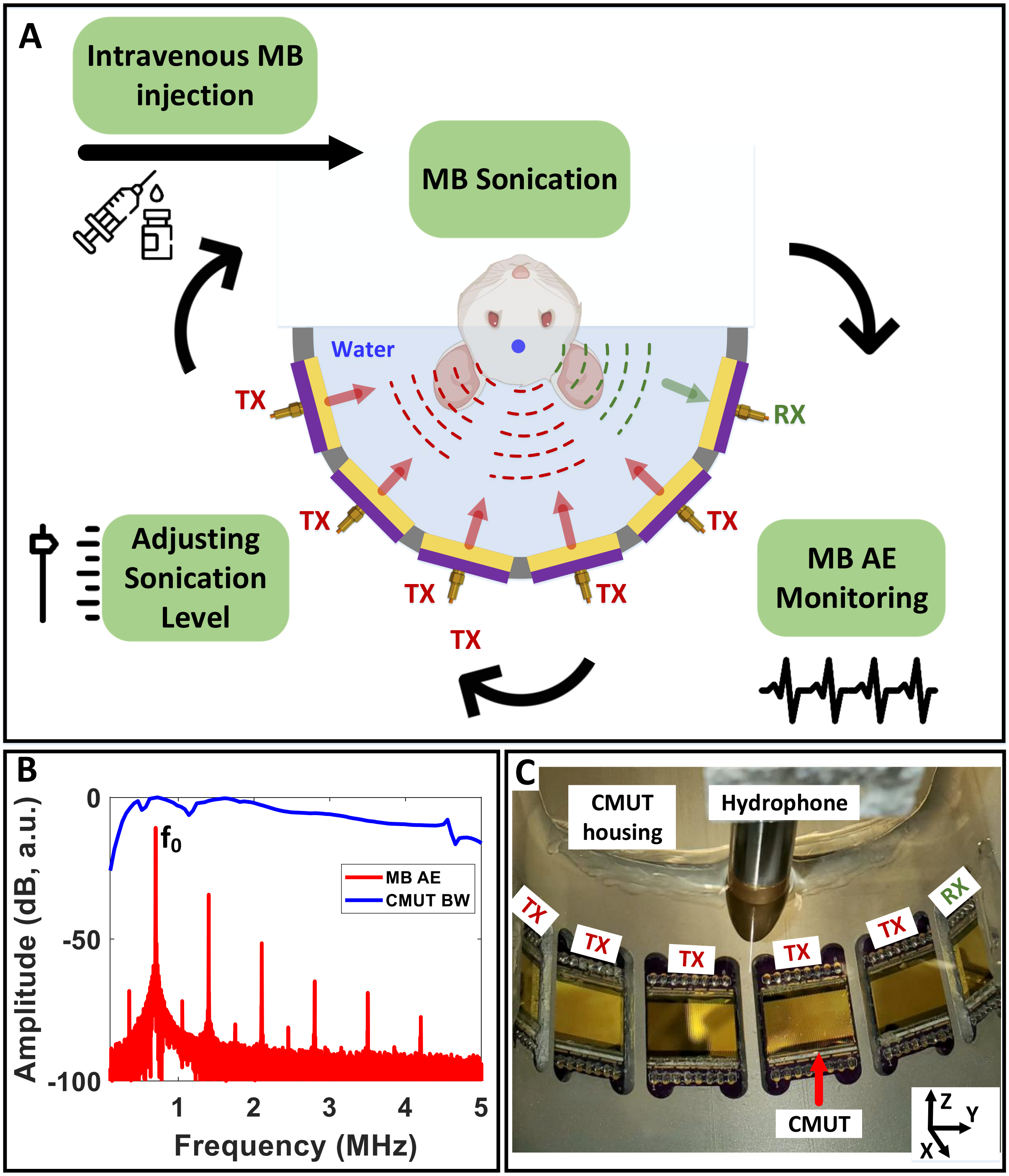

**Figure 1. System overview and acoustic characterization of the CMUT-based BBB-opening platform**

(A) Schematic of an ideal MB-assisted BBB opening workflow with a CMUT platform. The red dashed lines demonstrate the transmit (TX) waves and the green dashed line shows receive (RX) waves. (B) A representative MB acoustic emission spectrum (red) that shows the excitation frequency (f0) and the resulting sub/ultra and higher harmonics. Typical operational bandwidth of the CMUTs used in this study covering a broad frequency range for acoustic emission detection is shown in blue. (C) The 5 TX and 1 RX CMUT array implemented for the small animal MB-assisted BBB opening experiments. The geometric focus of the cylindrical array is 5 cm.

## RESULTS

To design a system for small-animal experimentation, we first simulated the transmit fields of the five-element CMUT TX arrays placed over a cylindrical structure using the acoustic modeling software Field II [31]. Note that due to the cylindrical structure (Fig. 1C), the array effectively generates a line focus where the size of the ellipsoid focal region in the elevation (x-direction) is mainly determined by the size of the CMUT element. Simulations confirmed that this configuration can achieve a tight focus of less than 2 mm in the azimuth direction (y-direction), focusing on a single side of a rat brain (Fig. 2A-B), so we transferred the CAD layout and fabricated the device housing. After assembling and sealing the device, we mapped the acoustic field with a calibrated hydrophone and compared it to the simulated field distribution (Fig. 2A-D). The close agreement in focal location, shape, and -6 dB beamwidth (Fig. 2E-G) validates the acoustic model and confirms proper alignment of the CMUT elements within the housing.

Subsequently, we assessed the ability to suppress CMUT-induced nonlinearities and enhance sensitivity to MB acoustic emissions using PI excitation and processing as a special case of phase modulation method [29, 35]. This method adds the received RF signals in response to two consecutive phase-inverted tone bursts (0° and 180°) and cancels odd-order harmonics originating from the transducer (Fig. 2H). Unlike our previous work that used a CMUT transmitter with a PVDF hydrophone receiver [30], the present study employs PI method for CMUTs in both transmission and reception, enabling suppression of receiver nonlinearities and a direct evaluation of an integrated all-CMUT based system. We first validated PI for CMUT nonlinearity suppression in degassed and deionized water using a 20-cycle, 700-kHz, sinewave burst excitation (Fig. 2I). Relative to single-phase excitation, PI reduced the fundamental component by 58 dB and the third harmonic by 36 dB in our CMUT system, confirming effective mitigation of device-generated nonlinearity. Although BBB-opening applications primarily operate below 4 MHz, we show spectra up to 6 MHz to demonstrate that this suppression behavior persists across the CMUT operational bandwidth under ideal background (no-MB) conditions. To validate enhanced isolation of MB acoustic emissions, we positioned a tube phantom within the array's focal region by maximizing the pulse-echo signal using five CMUT elements while the tube was filled with air. After the alignment, we filled the tube with degassed, deionized water and assessed the PI technique using two consecutive phase-inverted 7000-cycle (10 msec), 700 kHz, 170 kPa PNP excitation pulses with and without MBs flowing through the tube. The corresponding frequency spectra are shown in Fig. 2J. The deionized water (no-MB) condition showed no detectable subharmonic or ultraharmonic components, whereas the introduction of MBs lead to an increase of a 20 dB in the subharmonic, 15 dB in the first ultraharmonic, and 3 dB in the second ultraharmonic bands relative to the baseline. In addition, we observed 3 dB amplitude increase at the fundamental and 6 dB amplitude increase at the third harmonic level. After PI processing, the subharmonic and ultraharmonic levels remained constant, while the signal-to-noise ratio (SNR) at the fundamental and third-harmonic frequencies reached 37 dB and 13 dB, respectively, indicating the potential for isolating nonlinear fundamental signal detection generated by the MBs was earlier adapted by contrast-enhanced ultrasound imaging [32]. The zoomed spectra at the third harmonics, with a ±0.2 MHz range, in Fig. 2K highlight cancellation effects similar to Fig. 2I.

Together, these results demonstrate that the CMUT array produces sufficient acoustic pressure to activate MBs, possesses the bandwidth necessary to detect broadband acoustic emissions, and achieves harmonic suppression and sensitivity levels adequate for monitoring MB dynamics with high SNR- supporting its potential for integrated, closed-loop BBB-opening applications.

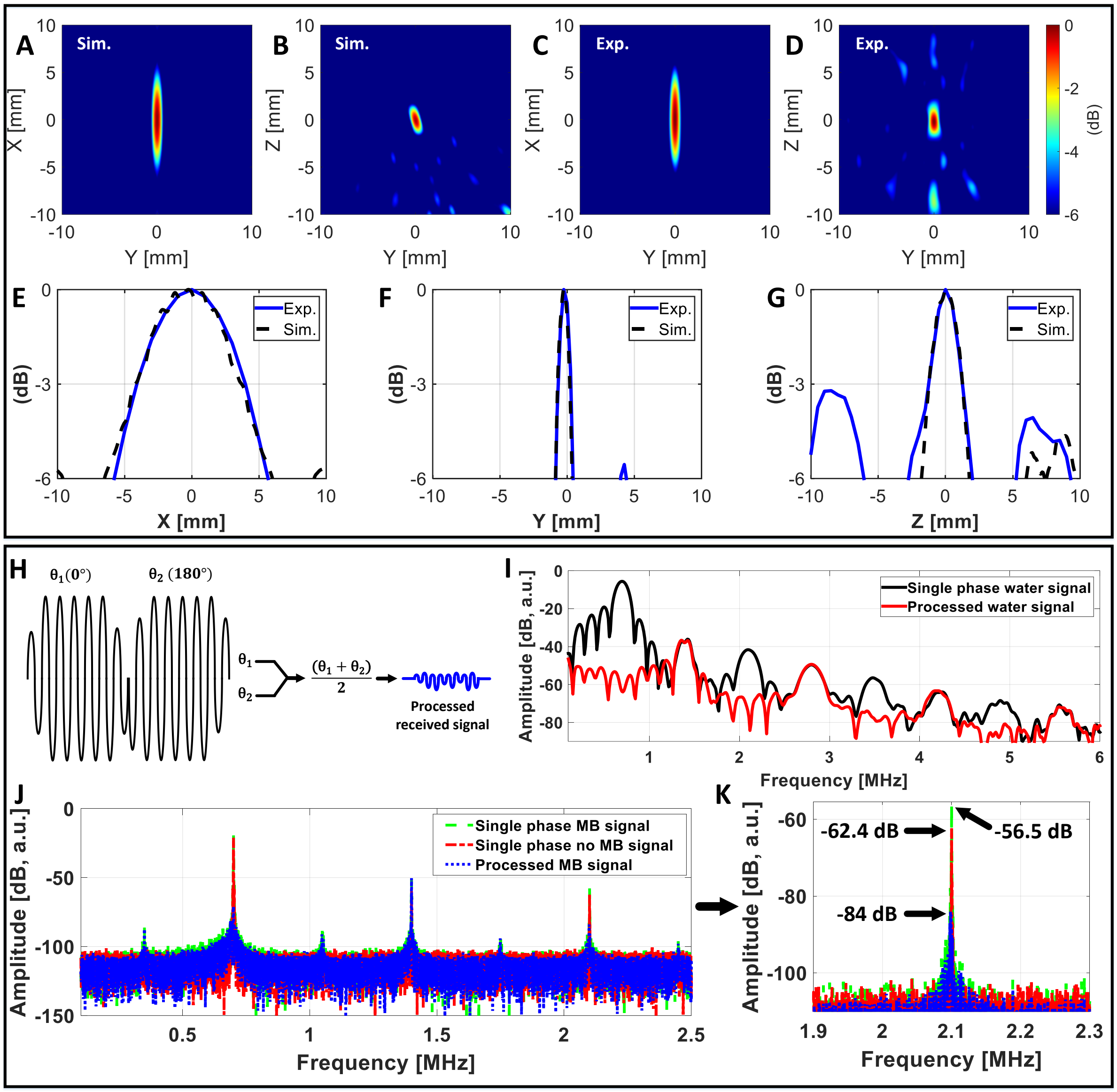

**Figure 2. Acoustic field validation and PI processing using the CMUT-based platform**

(A-B) Simulated normalized acoustic pressure fields of the five-element CMUT arrays in the lateral (XY) and axial (YZ) planes. (C-D) Experimentally measured normalized pressure fields acquired with a calibrated hydrophone at corresponding planes. (E-G) One-dimensional pressure profiles along the X, Y, and Z axes comparing simulation (black) and experimental (blue) results. (H) Schematic of the PI processing approach. Received signals from the CMUTs in response to two consecutive phase-inverted (0° and 180°) transmit signals in a single excitation, which are combined to cancel CMUT-generated nonlinearities while preserving MB-induced emissions at odd harmonics. (I) Water-tank validation using two consecutive, 20-cycle, 700 kHz excitation pulses. The single-phase water signal (black) and PI-processed signal (red) demonstrate the PI concept. (J) The received signal spectra in the case of an in vitro tube-phantom experiment with circulating MBs (700 kHz, 7000 cycles (10 msec)). (K) Zoomed in spectra around the third harmonic with a ±0.2 MHz range, indicating the suppression.

### *In vivo BBB opening with MRI confirmation and broadband Acoustic Emission monitoring using the CMUT platform*

Following the MB filled tube characterization, we evaluated the CMUT tFUS system in vivo during BBB-opening experiments in three rats (Fig. 3A). Each animal underwent two bilateral sonication conditions. The right hemisphere at high pressure (~800 kPa PNP in free field) and left hemisphere at low pressure (~400 kPa PNP in free field), to assess pressure-dependent acoustic emissions and BBB permeability. After anesthesia, we positioned the head at the acoustic focus and initiated sonication with the 5 TX elements and started recording of the acoustic emissions with the CMUT Rx element, 10 seconds before intravenous MB infusion. The sonication signal was 10ms long with 1s period, same as the previous experiments. Frequency-domain analysis of the recorded acoustic emission signals revealed strong nonlinear activity (Fig. 3B). The subharmonic level increased by 35 dB, the first and second ultraharmonic levels by 15 dB, and 12 dB, respectively, with 1 dB change at the fundamental and 4 dB change at third harmonic level relative to baseline noise. These results confirmed the CMUT array's ability to generate sufficient pressure at focus and to detect key spectral markers of MB dynamics during BBB opening.

To further analyze the MB activity with conventional processing in temporal domain, we tracked acoustic emission amplitudes over time for the fundamental, third-harmonic, and first-ultraharmonic components. The results are plotted here in dB relative to the average signal levels before MB infusion with the third-harmonic components corrected for linear reflectivity (Figs. 3C-E). These normalized outputs are relevant to closed-loop control as pressure differences with MB introduction are used to determine controller output [13, 33]. The high-pressure sonication produced a clear increase in the fundamental and 3rd harmonic bands with the introduction of the MBs and subsequent reduction as the MBs started to be cleared out. In contrast, the low pressure sonications showed variable responses. At both pressures the first ultraharmonic component exhibited inconsistent behavior.

Next, we processed the recorded acoustic emissions using PI processing and plotted the results at the same frequency bands in dB relative to the level of the processed signals before MB introduction to provide a similar reference, i.e. 0dB corresponds to average PI processed signal amplitude during the 10s before MB injection (Fig. 3F-H). As a specific application of phase modulation, PI selectively enhances nonlinear signals at odd harmonics. Consequently, for the high-pressure case the fundamental signal showed 20dB increase with MB introduction as compared to less than a dB with conventional processing. Importantly, under low pressure excitation a clear rise-fall trend in the nonlinear fundamental was observed by 13 dB increase, which was not visible with conventional processing. The third harmonic level showed 10 dB increase with MBs under high pressure compared to baseline, while low pressure remained consistently lower and variable. Likewise, ultraharmonic signals remained intermittent in both high- and low-pressure cases. Together these findings indicate that PI during sonication can improve the detection of pressure-dependent acoustic emission signatures, particularly in the fundamental and third-harmonic bands, while preserving subharmonic and ultraharmonic integrity.

Post-treatment T1-weighted and dynamic contrast-enhanced MRI (Bruker 7T) revealed localized, pressure-dependent contrast enhancement confined to targeted regions (Figs. 3I-L). Voxel wise $K_{trans}$ maps quantified permeability, having the following mean values across each region of interest: 0.0852 $min^{-1}$ (high-pressure hemisphere), 0.0352 $min^{-1}$ (low-pressure hemisphere), and 0.01 $min^{-1}$ (unsonicated control), confirming pressure-dependent BBB modulation consistent with predicted focal patterns (Fig. 3I-L).

Collectively, these results demonstrate that the CMUT-based system can promote and monitor localized BBB opening in rodents. Moreover, peak level of PI-based temporal MB acoustic emissions (including fundamental and third harmonic) strongly correlate with MRI-measured permeability, supporting the potential of broadband acoustic emission feedback for real-time control of ultrasound-mediated BBB opening.

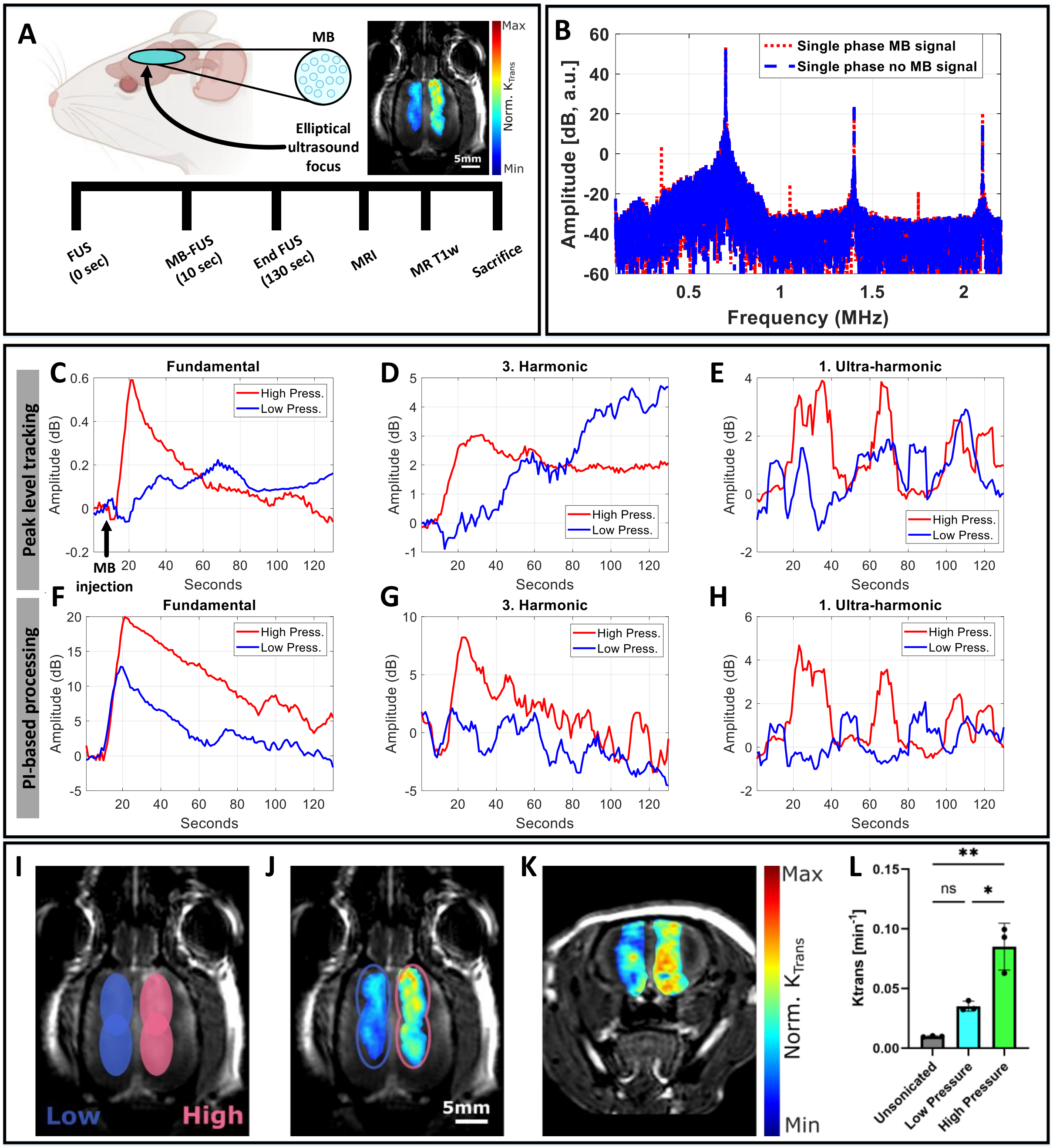

**Figure 3. In vivo validation of pressure-dependent BBB opening and Acoustic Emission monitoring using the CMUT-based system**

(A) Experimental timeline for BBB-opening protocol in rats and schematic of the in vivo sonication window showing the experimentally measured elliptical ultrasound focal region overlaid on the rat brain. (B) Frequency spectra of acoustic emissions recorded before (blue) and after (red) MB injection. (C-E) Temporal evolution of acoustic emission amplitudes normalized to average amplitude before MB injection for the fundamental, third-harmonic, and first-ultraharmonic components using peak level tracking. (F-H) Corresponding PI-processed tracking results. (I-K) Post-sonication T1-weighted MR images showing spatially localized BBB opening. (I) Regions of interest (ROIs) defined for high- and low-pressure sonications. (J) Normalized $K_{trans}$ overlay on the coronal image showing pressure-dependent permeability. (K) Axial T1-weighted image showing bilateral, spatially confined BBB disruption. (L) Quantified permeability ($K_{trans}$) across conditions.

## DISCUSSION

In this study, we designed and evaluated a CMUT-based tFUS platform that both delivers therapeutic ultrasound and monitors MB acoustic emissions across a broad frequency range. We showed that CMUT-generated nonlinearities can be substantially reduced using PI excitation under *in vivo (i.e., rodents)* conditions. Beyond validation of BBB opening, we characterized the temporal evolution of MB acoustic emissions and compared conventional peak-level tracking with PI-based processing, which markedly improved the dynamic range. Although acoustic emission data was acquired and analyzed in a post-processing setting in this work, future efforts will implement real-time processing and closed-loop feedback to regulate sonication parameters during BBB opening.

Beyond increasing the dynamic range, our findings indicate that the nonlinear fundamental component itself may serve as a useful metric to track MB kinetics and the extent of BBB opening, an approach that, to our knowledge, has not been systematically investigated, as most prior work has focused on subharmonics, higher-order harmonics, and ultraharmonics. PI processing effectively normalizes the baseline by suppressing setup-dependent variation, so that before the MB arrival, the fundamental levels at low and high pressure are comparable; after MB arrival, the fundamental peak clearly separates the two pressure conditions, particularly in the high-pressure case, where a more pronounced rise-fall pattern is observed. This behavior suggests that the PI-processed fundamental could be incorporated into feedback control schemes, where tracking of MB kinetics is important [13, 34].

In terms of the third-harmonic response, which is one of the commonly used spectral components for monitoring and controlling MB-assisted BBB opening, the high pressure case showed consistent behavior both with conventional and PI processing, whereas the third harmonic acoustic emissions for the low pressure excitation was inconsistent, albeit showing a more expected trend for the PI processing. The ultraharmonic signals are not impacted by CMUT nonlinearity as CMUTs do not generate ultraharmonics [35]. Therefore, as expected, conventional and PI processed ultraharmonic signals are similar to each other. Overall, these results indicate that CMUT receivers can provide acoustic emissions signals with high fidelity, especially with PI processing.

MRI findings (Figs. 3J-K) confirmed that BBB opening was spatially confined to the targeted regions, and permeability mapping ($K_{trans}$) quantified a clear pressure dependence. High-pressure sonications produced greater permeability than low-pressure ones, and both exceeded the unsonicated control. These results align with the focal patterns predicted by acoustic field simulations (Figs. 2C-D). However, while permeability correlated with sonication pressure, the relationship between specific acoustic emission features, such as the nonlinear fundamental amplitude (Fig. 3F), and the magnitude of BBB opening remains to be fully established. Establishing such correlations between spectral features (e.g., fundamental amplitude, third-harmonic growth, ultraharmonic bursting) and quantitative permeability metrics will be crucial for developing comprehensive closed-loop control strategies.

Several technical challenges remain in the current system. Transmit and receive functions are implemented with separate CMUT elements, but each CMUT array has 64 channels. Leveraging these channels for electronic beamforming and delay-and-sum reception could enable focal steering, spatial filtering of acoustic emission signals, and improved localization of cavitation activity, allowing multi-site sonication without physical repositioning. On the receive side, the system currently uses a general-purpose amplifier rather than a CMUT-optimized low-noise transimpedance front end. Co-designing CMUTs with dedicated readout electronics would improve sensitivity and dynamic range. In parallel, advanced excitation and decoding schemes, such as polyphase modulation and adaptive filtering, could further suppress device-originated nonlinearities while preserving bubble-derived emissions. Additionally, proof-of-conceptstudy was conducted in a small-animal cohort with a limited number of sonication sites per animal; expanding the sample size and correlating MB acoustic emission spectra with MRI and histology based safety metrics will strengthen translational relevance.

Together our results indicated that, integrating real-time PI processing, CMUT-specific electronics, multi-channel beamforming, and controlled multi-site sonication can advance the current all-CMUT platform from preclinical feasibility toward a human-scale, feedback-controlled BBB-opening system.

## METHODS

### *tFUS Driving and Signal Acquisition Setup*

In this study, a 5-cm geometrically focused six-element CMUT (CM5, Philips Innovations) assembly was used, consisting of five TX elements for focused ultrasound delivery and one RX element for passive detection of MB AEs. The elements were mounted in a custom-designed 3D-printed cylindrical housing whose acoustic characterization at 400 kHz has been reported previously [30]. Similarly, TX elements were driven by an arbitrary waveform generator (Model 33600A, Keysight Technologies, Santa Clara, CA, USA), which produced a 700 kHz excitation consisting of two consecutive 0° and 180° phase with 20 and 3500 number of cycles per drive signal at a pulse repetition frequency (PRF) of 1 Hz. A DC bias of 100 V, supplied by a power supply (Model PS310, Stanford Research Systems, Sunnyvale, CA, USA), was applied to operate the CMUTs in the collapsed mode. The RX CMUT element was connected to a broadband pulser/receiver unit (Model 5072PR, Olympus, Waltham, MA, USA) configured with a 3 dB gain. The output of the amplifier was digitized using a 14-bit oscilloscope (PicoScope 5442D, Pico Technology Ltd., St. Neots, UK) and transferred to a computer for subsequent post-processing. The system was synchronized via trigger outputs from the waveform generator to ensure precise timing between transmit and receive events.

### *Animal Preparation and Anesthesia Protocol*

All animal experiments were conducted in compliance with the Public Health Service Policy on Humane Care and Use of Laboratory Animals and were approved by the Institutional Animal Care and Use Committee (IACUC) of the Georgia Institute of Technology.

Healthy male Sprague-Dawley rats (approximately 180 g, Charles River Laboratory, total n = 2) were used in this study. Anesthesia was induced via mask induction of isofluorane (continuously 1.5% concentration). The fur on the scalp was shaved and further removed with a depilatory cream to ensure optimal acoustic coupling. The tail vein was catheterized to enable intravenous administration of microbubbles and MR contrast agents during and after experiments, respectively.

### *In Vivo Sonication and Acoustic Emission Monitoring*

For *in vivo* experiments, anesthetized rats were positioned supine in a custom-built holder that was mounted on a three-axis micromanipulator. The head was aligned so that the expected acoustic focus of the CMUT array coincided with the desired intracranial target. Prior to each experiment, the focal location was verified using a 3D-printed alignment tip inserted into the CMUT housing to provide a visual/mechanical reference for targeting. Each animal received two sonications: one targeting a “high-pressure” site (~800 kPa free field peak negative pressure (PNP)) and one targeting a “low-pressure” site (~400 kPa free field PNP). These sites were placed in opposite hemispheres to evaluate pressure-dependent BBB permeability within the same subject. Sonications were performed at a center frequency of 700 kHz using 7000 cycles (10 ms) bursts at a PRF of 1 Hz, for a total exposure duration of approximately 130 s per site. The first 10 s of sonication were recorded prior to MB injection to establish a baseline (“no-MB” condition), followed by 120 s of sonication acquired during MB circulation after intravenous bolus injection.

MBs were administered as an intravenous bolus via a tail-vein catheter immediately prior to and during sonication. For experiments, an in-house Definity-like lipid MB formulation was used due to limited commercial supply. Throughout sonication, AEs from the focal region were passively acquired using the receive CMUT element operating as a PCD.

acoustic emission analysis was carried out offline. For PI processing, the time-domain receive signal was digitized, the two phase signals were separated, windowed with a Hann window, and added in time domain. The resulting time data was converted to the frequency domain using a fast Fourier transform (FFT) in MATLAB (MathWorks). Cavitation-related enhancement at specific frequency components (subharmonic, harmonic, ultraharmonic) was then quantified as the relative increase in spectral level.

### *MRI Acquisition and Permeability Analysis*

To quantify changes in BBB permeability after sonication, immediately after the ultrasound exposure DCE-MRI was carried out using a 7 T small-animal MRI scanner (Pharmascan 7T, Bruker). Following the final ultrasound exposure, DCE imaging was performed with a gradient-echo sequence (echo time: 2.5 ms; repetition time: 1019.6 ms; flip angle: 30°; field of view: 35 × 35 $mm^2$). The main quantitative parameter extracted was the volume transfer constant ($K_{trans}$), which characterizes the rate of contrast-agent exchange between blood plasma and the extravascular-extracellular compartment and serves as a measure of BBB permeability. Prior to contrast injection, a baseline T1 map was generated by acquiring a series of pre-contrast images at multiple flip angles (2°, 5°, 10°, 15°, 20°, and 30°). These datasets were used to compute voxelwise pre-contrast T1 values, which were then applied to convert DCE signal variations over time into concentration-time profiles for subsequent kinetic modeling.

Following acquisition of the baseline images, DCE scanning was initiated concurrently with an intravenous bolus injection of a gadolinium-based contrast agent (ProHance; 2.0 mL/kg, equivalent to 0.4 mL in this study). Sequential images were continuously acquired to monitor the contrast-agent uptake and clearance dynamics within the targeted brain regions. Image analysis was carried out using Horos software equipped with the DCE Tool plugin (Kyung Sung, Los Angeles, CA). For each defined region of interest, $K_{trans}$ was derived by fitting the calculated tissue concentration-time profiles to a standard pharmacokinetic model implemented in the plugin. The arterial input function (AIF) necessary for this modeling was estimated using the Fritz-Hansen method provided within the same software [36]. The resulting $K_{trans}$ parametric maps enabled quantitative comparison of BBB permeability across high-pressure, low-pressure, and control (unsonicated) regions. All $K_{trans}$ results were obtained from the mean value across each individual region of interest and averaged across all three imaged depths (each 1 mm apart).